# The large-scale gravitational bias from the quasi-linear regime

**F. Bernardeau**

Service de Physique Théorique*, CE de Saclay, F-91191 Gif-sur-Yvette Cedex, France

**Abstract.** It is known that in gravitational instability scenarios the nonlinear dynamics induces non-Gaussian features in cosmological density fields that can be investigated with perturbation theory. Here, I derive the expression of the joint moments of cosmological density fields taken at two different locations. The results are valid when the density fields are filtered with a top-hat filter window function, and when the distance between the two cells is large compared to the smoothing length. In particular I show that it is possible to get the generating function of the coefficients $C_{p,q}$ defined by $\left\langle\delta^p(\boldsymbol{x}_1)\delta^q(\boldsymbol{x}_2)\right\rangle_c = C_{p,q}\left\langle\delta^2(\boldsymbol{x})\right\rangle^{p+q-2}\left\langle\delta(\boldsymbol{x}_1)\delta(\boldsymbol{x}_2)\right\rangle$ where $\delta(\boldsymbol{x})$ is the local smoothed density field. It is then possible to reconstruct the joint density probability distribution function (PDF), generalizing for two points what has been obtained previously for the one-point density PDF. I discuss the validity of the large separation approximation in an explicit numerical Monte Carlo integration of the $C_{2,1}$ parameter as a function of $|\boldsymbol{x}_1 - \boldsymbol{x}_2|$.

A straightforward application is the calculation of the large-scale "bias" properties of the over-dense (or under-dense) regions. The properties and the shape of the bias function are presented in details and successfully compared with numerical results obtained in an N-body simulation with CDM initial conditions.

## 1. Introduction

Recently a lot of progress has been made in the study of the nonlinear features induced by the gravitational dynamics in an expanding Universe. In particular it has been shown that it is possible to apply the techniques of perturbation theory (PT) to derive statistical properties of the density field at large-scale. For instance it is possible to use PT to calculate the leading order of the low order reduced moments of the local density, (Peebles 1980, Fry 1984). Refining these calculations Goroff et al. (1986) showed that the unavoidable filtering of the cosmological density fields affects the results. They calculated numerically those effects in a particular case. More recently Juszkiewicz (1993), Bernardeau (1994b), Lokas et al. (1995) investigated by means of analytic calculations these smoothing effects either for a top-hat or a Gaussian window function. And finally the initial results obtained for the third and fourth cumulants of the one-point density distribution function, have been extended to the whole series of the cumulants by Bernardeau (1992) when the smoothing effects are neglected and Bernardeau (1994c) for top-hat filtering. These results are found to be in remarkable agreement with results of numerical simulations (Juszkiewicz et al. 1995, Bernardeau 1994b,c, Baugh, Gaztañaga & Efstathiou 1995, Gaztañaga & Baugh 1995, Lokas et al. 1995). So far, however, the properties of the moments of the two-point density PDF have never been investigated.

This is however a quantity of various interests. Its most direct application would probably be for a proper estimation of the errors due to finite sampling for the measurements of the statistical properties of the one-point density field. Indeed, when the density is measured in a finite sample at different locations, these measures are not independent and the estimation of the errors should take into account these correlations. The two point density PDF is expected to provide the most dominant part of these correlations (Colombi, Bouchet & Schaeffer 1995, Szapudi & Colombi 1995).

The derivation of the joint-density PDF is also of crucial interest for the study of the properties of a gravitational fluid. In particular it can give insights on how the dense spots of the field are correlated together, compared to the global matter correlation function. This is thus related to the general problem of the bias in large-

* Laboratoire de la Direction des Sciences de la Matière du Commissariat à l'Energie Atomique

scale structure formation. The problem that shall eventually be addressed here is the estimation of the gravity induced bias in the quasi-linear regime.

The most commonly considered bias factor is the bias $b$ associated to the galaxies. In general the problem of the bias emerges because the mass distribution of the universe cannot be observed directly and thus has to be estimated with the apparent distribution of tracers. However, these tracers, galaxies, (or galaxy clusters, super-clusters) may not be reliable tracers of the underlying mass distribution so that their density fluctuations do not reproduce the underlying mass density fluctuations. From the observations of the density fluctuations of the clusters and the galaxies, it is already clear that there is a relative bias between these objects. This concern led people to introduce a bias parameter $b$ relying the distribution of tracers to the mass distribution. In its simplest form it is assumed that the local (smoothed) over-density of objects, $\delta_{\rm gal.}$, is proportional to the local mass over-density,

$$\delta_{\rm gal.} = b\ \delta_{\rm mass}. \tag{1}$$

The physical processes that can possibly affect the value of $b$ are numerous, but at very large scale it is likely that only gravity can substantially affect the correlation of large mass concentrations. Neglecting, at first sight, the problem of identifying the actual astrophysical objects in the evolved final density field, we are led to a purely gravitational physics problem, that is the calculation of the correlation of dense spots in a self-gravitating fluid.

There have been attempts to solve this latter problem. For instance Doroshkevich & Shandarin (1978a, b) and Bardeen et al. (1986) derived the bias property of the peaks in a Gaussian density field. This idea had been suggested by Kaiser (1984) to explain the relatively strong clustering of the Abell clusters. This gave a good insight of the possible selection effects in a random field. This result, however, did not take into account the subsequent nonlinear dynamics which is likely to affect the correlation properties of the peaks, such as their peculiar displacement, or more worrying, the possible merging of halos. In an attempt to overcome these difficulties and starting with a different point of view - a phenomenological description of the matter $p$-point correlation functions in the final nonlinear field - Schaeffer (1985) and more recently Bernardeau & Schaeffer (1992), extending a work of Balian & Schaeffer (1989), estimated the expected bias of the observed objects induced by gravity. Their study, however, lacked definitive quantitative prediction due to the absence of a complete and reliable description of the matter correlation properties in the nonlinear regime. In this paper I would like to reconsider some of these calculations but with precise quantitative results obtained in the quasi-linear regime. This is certainly not a complete derivation of the gravity induced peak correlations in a cosmological density field. In particular the positions of the cells are not constrained to be local maxima of the smoothed density field. The aim of my study is simply to derive the joint density PDF at two different distant locations. In practice these results are applicable only to the largest mass concentrations observed in the Universe (the smoothing length should be larger than the correlation length) and for large separation. But it can give insights in the gravity induced large-scale biases.

The paper is divided as followed. In §2 I present the general properties of the cumulants for joint densities in the quasi-linear regime and in the large-separation approximation. The quantitative derivations are given in §3. In §4 I discuss the validity of the large-separation approximation. Eventually, §5 is devoted to comparisons with numerical simulations and §6 to a discussion of the possible astrophysical implications.

## 2. General properties of the cumulants of the two-point density PDF

I assume here that the cosmological density field evolved from Gaussian initial fluctuations following the gravitational dynamics of a pressure-less fluid in an expanding universe. Moreover the cosmological over-density field $\delta(\boldsymbol{x})(\equiv \rho(\boldsymbol{x})\overline{\rho} - 1)$ is assumed to be filtered by a given window function so that the local rms fluctuations are small. Although the general properties discussed in this section are valid for any window function, most of the analytical results presented afterwards are valid for a top-hat window function only. The numerical application will be done for an Einstein-de Sitter Universe.

I assume here that the reader knows how cumulants, noted in the following $\langle . \rangle_c$, are defined, otherwise see for instance Balian & Schaeffer (1989).

### *2.1. Notations and diagrammatic representations*

The Fourier transform, $\delta(\boldsymbol{k})$, of the initial density field is defined by

$$\delta(\boldsymbol{x}) = \int \mathrm{d}^3\boldsymbol{k}\,\delta(\boldsymbol{k})\exp(\mathrm{i}\boldsymbol{k}\cdot\boldsymbol{x}), \tag{2}$$

where $\delta(\boldsymbol{k})$ are random Gaussian variables. The power spectrum is given by

$$\big\langle\delta(\boldsymbol{k})\delta(\boldsymbol{k}')\big\rangle = \delta_{\rm Dirac}(\boldsymbol{k}+\boldsymbol{k}')\,P(k). \tag{3}$$

The hypothesis of Gaussian variables is obviously of crucial interest for the derivation of the subsequent statistical properties of the density field. It implies in particular that the ensemble average of a product of an odd number of variables $\delta(\boldsymbol{k})$ is zero and that the ensemble average of a product of an even number is obtained by summing all the possible different products of cross-correlation when the $\delta(\boldsymbol{k})$ factors are associated by pairs.

In the quasi-linear regime the evolution of the density field is governed by the continuity, Euler and Poisson equations. Let me consider and expansion of the growing modes of the density field with the initial fluctuations field,

$$\delta(\boldsymbol{x}) = \delta^{(1)}(\boldsymbol{x}) + \delta^{(2)}(\boldsymbol{x}) + \delta^{(3)}(\boldsymbol{x}) + \dots \quad (4)$$

where $\delta^{(1)}$ is proportional to the initial fluctuations, $\delta^{(2)}$ quadratic, etc.. Following Goroff et al. (1986) one can write the expression of the density field at the order $p$, smoothed at the scale $R$,

$$\delta^{(p)} = \int \mathrm{d}^3\boldsymbol{k}_1 \dots \int \mathrm{d}^3\boldsymbol{k}_p \delta(\boldsymbol{k}_1)\dots\delta(\boldsymbol{k}_p) \times N_p(\boldsymbol{k}_1,\dots,\boldsymbol{k}_p)\, W\left[|\boldsymbol{k}_1+\dots\boldsymbol{k}_p|\; R\right] D^p(t) \quad (5)$$

where $D(t)$ is the time dependence of the linear growing mode, $N_p$ are homogeneous function of the wave vectors, $W$ is the Fourier transform of the window function applied to the field. (Note that this form is valid only for an Einstein-de Sitter Universe but admits simple generalizations).

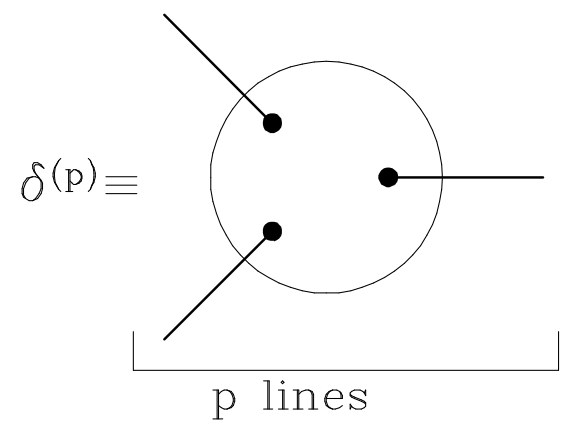

**Fig. 1.** Diagrammatic representation of the term $\delta^{(p)}(\boldsymbol{x})$. Each dot stands for a factor $\delta(\boldsymbol{k})$.

In Fig. 1 I present a diagrammatic representation of the term $\delta^{(p)}(\boldsymbol{x})$. Each point appearing in the diagram stands for a factor $\delta(\boldsymbol{k})$; the lines for $\exp(\mathrm{i}\boldsymbol{k}\cdot\boldsymbol{x})$ and the hypothesis of Gaussian hypothesis implies that the ensemble average of an arbitrary number of such quantities should be calculated by associating the points by pairs (as in Figs. 2 and 3).

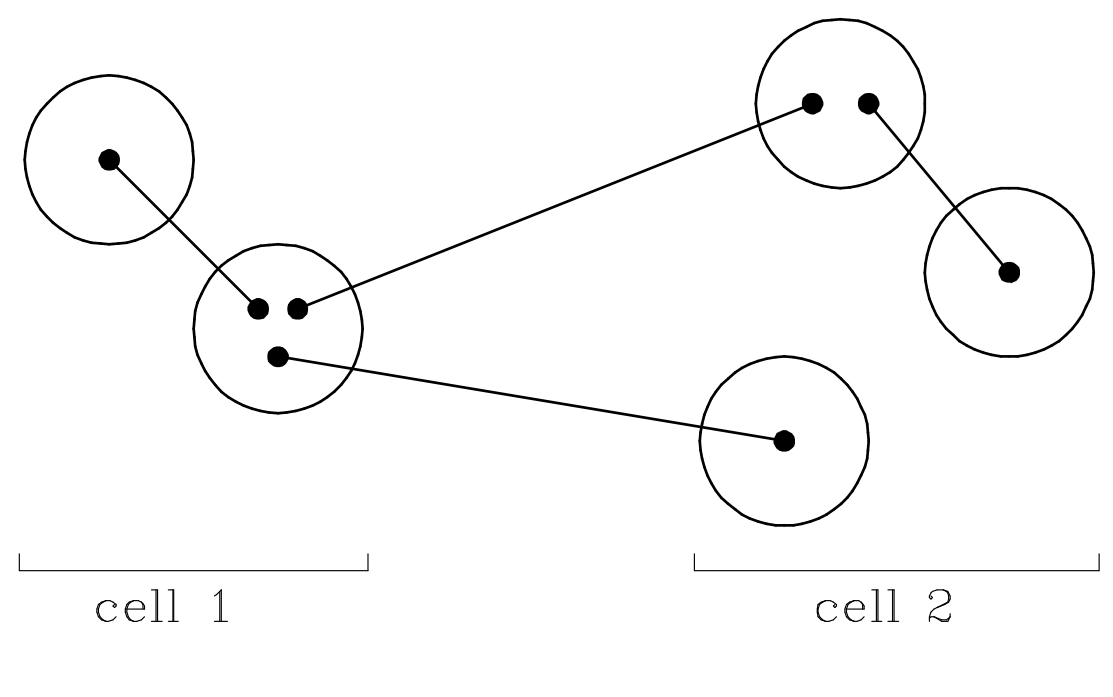

**Fig. 2.** Example of a term contributing to $\left\langle\delta^2(\boldsymbol{x}_1)\delta^3(\boldsymbol{x}_2)\right\rangle_c$. The circles at the left side of the figure are at the position $\boldsymbol{x}_1$ and the ones at the right side are at the position $\boldsymbol{x}_2$.

In the following I am interested in the calculation of the dominant order of the cumulants of the kind $\left\langle\delta^p(\boldsymbol{x}_1)\delta^q(\boldsymbol{x}_2)\right\rangle_c$, where $\delta(\boldsymbol{x})$ is the smoothed local density at the position $\boldsymbol{x}$. Such cumulants at their leading order in the quasi-linear regime are given by an expression of the form

$$\left\langle\delta^p(\boldsymbol{x}_1)\delta^q(\boldsymbol{x}_2)\right\rangle_c \approx \sum_{\text{decompositions}} \left\langle \prod_{i=1}^{p} \delta^{(p_i)}(\boldsymbol{x}_1) \prod_{i=1}^{q} \delta^{(q_i)}(\boldsymbol{x}_2)\right\rangle_c \quad (6)$$

where the decompositions are made in such a way that the integers $p_i$ and $q_j$ verify

$$\sum_{i=1}^{p} p_i + \sum_{i=1}^{q} q_i = 2(p+q) - 2. \quad (7)$$

This constraint ensures that the number of lines connecting the factors is large enough for the term to be connected (terms in which the sum (7) is less than $2(p+q)-2$ do not contribute to the cumulants), but small enough that there are no extra lines that would lead to a negligible term in the quasi-linear regime. In Fig. 2 I present the diagrammatic representation of such a term contributing to the expression of $\left\langle\delta^2(\boldsymbol{x}_1)\delta^3(\boldsymbol{x}_2)\right\rangle_c$. This particular term enters in the determination of $\left\langle\delta^{(1)}(\boldsymbol{x}_1)\delta^{(3)}(\boldsymbol{x}_1)\left[\delta^{(1)}(\boldsymbol{x}_2)\right]^2\delta^{(2)}(\boldsymbol{x}_2)\right\rangle_c$. It is of interest to note that for $q=0$ (or $p=0$) the constraint (7) reduces to

$$\sum_{i=1}^{p} p_i = 2p - 2, \quad (8)$$

and it corresponds to the calculations of the cumulants of the one-point density PDF. The latter are then expected to behave like (Fry 1984, Goroff et al. 1986, Bernardeau 1992)

$$\left\langle\delta^p(\boldsymbol{x})\right\rangle_c = S_p \left\langle\delta^2(\boldsymbol{x})\right\rangle^{p-1}, \quad (9)$$

where the power $p-1$ is a direct consequence of the constraint (8). The coefficients $S_p$ have been calculated exactly for a top-hat window function (Bernardeau 1994c). The calculation presented in the following is a generalization of these results to joint cumulants.

### *2.2. The large separation approximation, factorization properties*

The general calculation of diagrams as in Figs. 2, 3 is unfortunately quite complicated and, as far as I know, cannot be done in general. However, the calculations can be carried on in a limiting case, when the separation between the cells is large. This limit can be understood in a diagrammatic point of view. The length of the lines joining two points in Fig. 2 is roughly proportional to the value of the two-point correlation function at the same scale. It implies that long lines (for hierarchical, CDM like, power spectrum) give a negligible contribution compared to short lines. As a result, terms in which the number of long lines (the ones that connect the cells together) is minimal dominate the expression of the joint cumulants. For instance in Fig. 3 I present a diagram whose contribution to the expression of $\left\langle\delta^2(\boldsymbol{x}_1)\delta^3(\boldsymbol{x}_2)\right\rangle_c$ is dominant compared to the one presented in Fig. 2 in the large separation limit. For two cells the number of lines joining the two cells should simply be one. The dominant contributions are now given by an expression similar to (6) but with

$$\sum_{i=1}^{p} p_i = 2p-1, \tag{10a}$$

and

$$\sum_{i=1}^{q} q_i = 2q-1. \tag{10b}$$

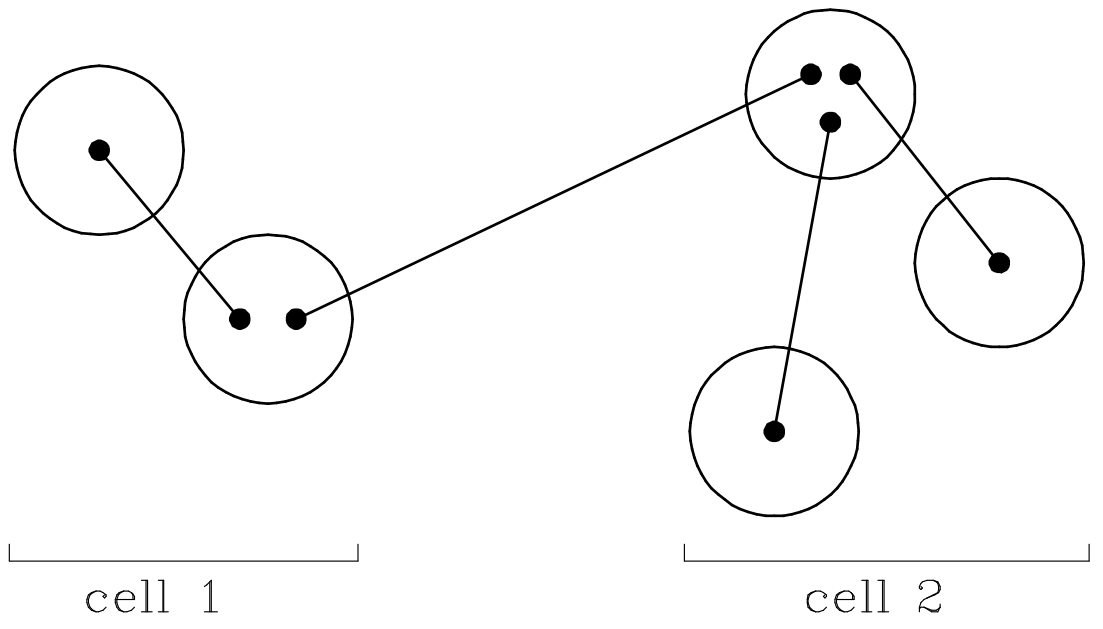

**Fig. 3.** Example of a term contributing to $\left\langle\delta^2(\boldsymbol{x}_1)\delta^3(\boldsymbol{x}_2)\right\rangle_c$. The circles at the left side of the figure are at the position $\boldsymbol{x}_1$ and the ones at the right side are at the position $\boldsymbol{x}_2$.

So, in the large separation limit, only a subset of diagrams have to be calculated. The general form of such diagrams is given by

$$\left\langle\delta^p(\boldsymbol{x}_1)\delta^q(\boldsymbol{x}_2)\right\rangle_c = \int \mathrm{d}^3\boldsymbol{k}_0\, G_p(\boldsymbol{k}_0,R)\, G_q(\boldsymbol{k}_0,R)\, P(k_0) \times \exp\left[\mathrm{i}\boldsymbol{k}_0\cdot(\boldsymbol{x}_1-\boldsymbol{x}_2)\right] \tag{11}$$

with

$$G_p(\boldsymbol{k}_0,R) = \int \mathrm{d}^3\boldsymbol{k}_1\dots\mathrm{d}^3\boldsymbol{k}_{p-1} \times \sum_{\text{permutations }\{s\}} \prod_{i=1}^{p} F_{p_i}(\boldsymbol{k}_{s_1},\dots,\boldsymbol{k}_{s_{p_i}}) \times W\left[|\boldsymbol{k}_{s_1}+\dots+\boldsymbol{k}_{s_{p_i}}|\,R\right]\; P(k_1)\dots P(k_{p-1}) \tag{12}$$

with $G_1(\boldsymbol{k}_0,R)=1$ and where $\boldsymbol{k}_{s_j}$ is either $\boldsymbol{k}_1$, $\boldsymbol{k}_{p-1}$ or $\boldsymbol{k}_0$. The determination of the integrals (11) is then reduced to the derivation of a single index series of integrals. The functions $G_p(\boldsymbol{k}_0,R)$, for dimensional reasons, depend on $k_0$ only through the combination $k_0R$. In the large separation limit the contributing value of $k_0$ in (12) will be of the order of $1/|\boldsymbol{x}_1-\boldsymbol{x}_2|$, so that the function $G(k_0,R)$ has to be evaluated only in the limit $k_0\to 0$. This makes the calculation quite simpler, since the cumulants (6) now read,

$$\begin{aligned}\left\langle\delta^p(\boldsymbol{x}_1)\delta^q(\boldsymbol{x}_2)\right\rangle_c &= G_p(0,R)\; G_q(0,R) \\ &\times \int \mathrm{d}^3\boldsymbol{k}_0 P(k_0)\, \exp\left[\mathrm{i}\boldsymbol{k}_0\cdot(\boldsymbol{x}_1-\boldsymbol{x}_2)\right] \\ &= G_p(0,R)\; G_q(0,R)\left\langle\delta(\boldsymbol{x}_1)\delta(\boldsymbol{x}_2)\right\rangle.\end{aligned} \tag{13}$$

I then define the coefficients $C_{p,q}$ by

$$\left\langle\delta^p(\boldsymbol{x}_1)\delta^q(\boldsymbol{x}_2)\right\rangle_c = C_{p,q}\left\langle\delta^2\right\rangle^{p+q-2}\left\langle\delta(\boldsymbol{x}_1)\delta(\boldsymbol{x}_2)\right\rangle. \tag{14}$$

It is then clear that the dominant term in the quasi-linear regime for the expression of $C_{p,q}$ is constant in the limit of large separation (i.e., independent of $d$). Moreover from the previous equation the coefficient $C_{p,q}$ is simply given by ,

$$C_{p,q} = \frac{G_p(0,R)}{\left\langle\delta^2\right\rangle^{p-1}}\,\frac{G_q(0,R)}{\left\langle\delta^2\right\rangle^{q-1}} = C_{p,1}\,C_{q,1}, \tag{15}$$

and is thus factorizable in $p$ and $q$. This is one of the key properties obtained from the large-separation limit approximation.

### *2.3. Examples of coefficients: $C_{2,1}$ and $C_{2,2}$*

In this section I derive the values of $C_{2,1}$ and $C_{2,2}$ in the large-scale limit for a top-hat window function. In such a case the window function in Fourier space is given by

$$W(k) = \frac{3}{k^3}\left[\sin(k) - k\cos(k)\right]. \tag{16}$$

Before doing the calculations I recall few properties of this window function, given by Bernardeau (1994a),

$$\frac{1}{4\pi}\int \mathrm{d}\Omega\, W\left[|\boldsymbol{k}_1+\boldsymbol{k}_2|\right]\left[1-\frac{(\boldsymbol{k}_1\cdot\boldsymbol{k}_2)^2}{k_1^2\,k_2^2}\right] = \frac{2}{3}\,W(k_1)\,W(k_2), \tag{17}$$

and

$$\frac{1}{4\pi}\int \mathrm{d}\Omega\, W\left[|\boldsymbol{k}_1+\boldsymbol{k}_2|\right]\left(1+\frac{\boldsymbol{k}_1\cdot\boldsymbol{k}_2}{k_1^2}\right) = W(k_1)\left[W(k_2)+\frac{1}{3}k_2\,W'(k_2)\right], \tag{18}$$

where the integration is made indifferently over the angular part of $\boldsymbol{k}_1$ or $\boldsymbol{k}_2$.

The expression of the coefficient $C_{2,1}$ is given by

$$\begin{aligned}C_{2,1} =& \frac{4}{\big\langle\delta(\boldsymbol{x}_1)\delta(\boldsymbol{x}_2)\big\rangle\big\langle\delta^2\big\rangle} \\ &\times\int \mathrm{d}^3\boldsymbol{k}_0\mathrm{d}^3\boldsymbol{k}_1\ P(k_0)\ P(k_1)\ \exp\left[\mathrm{i}\boldsymbol{k}_0\cdot(\boldsymbol{x}_1-\boldsymbol{x}_2)\right] \\ &\times W(k_0R)\ W\left[\vert\boldsymbol{k}_0+\boldsymbol{k}_1\vert\,R\right]W(k_1R)\ N_2(\boldsymbol{k}_0,\boldsymbol{k}_1) \\ &+\frac{2}{\big\langle\delta(\boldsymbol{x}_1)\delta(\boldsymbol{x}_2)\big\rangle\big\langle\delta^2\big\rangle} \\ &\times\int \mathrm{d}^3\boldsymbol{k}_0\mathrm{d}^3\boldsymbol{k}_1\ P(k_0)\ P(k_1) \\ &\times\exp\left[\mathrm{i}(\boldsymbol{k}_0+\boldsymbol{k}_1)\cdot(\boldsymbol{x}_1-\boldsymbol{x}_2)\right] \\ &\times W(k_0R)\ W\left[\vert\boldsymbol{k}_0+\boldsymbol{k}_1\vert\,R\right]W(k_1R)\ F_2(\boldsymbol{k}_0,\boldsymbol{k}_1)\end{aligned} \qquad (19)$$

with

$$N_2(\boldsymbol{k}_0,\boldsymbol{k}_1)=\frac{5}{7}+\frac{1}{2}\frac{\boldsymbol{k}_0\cdot\boldsymbol{k}_1}{k_0^2}+\frac{1}{2}\frac{\boldsymbol{k}_0\cdot\boldsymbol{k}_1}{k_1^2}+\frac{2}{7}\left(\frac{\boldsymbol{k}_0\cdot\boldsymbol{k}_1}{k_0\,k_1}\right)^2.$$

The second term appearing in (19) is in fact negligible since it involves the presence of two lines joining the cells at the positions $\boldsymbol{x}_1$ and $\boldsymbol{x}_2$. The first term of this expression can be calculated using the properties (17-18) (integrating first on the angular part of the wave vector $\boldsymbol{k}_1$). It leads to the expression,

$$C_{2,1}=\frac{68}{21}+\frac{1}{3}\frac{\mathrm{d}\log\big\langle\delta^2\big\rangle}{\mathrm{d}\log R}+\frac{1}{3}\frac{\mathrm{d}\log\big\langle\delta(\boldsymbol{x}_1)\delta(\boldsymbol{x}_2)\big\rangle}{\mathrm{d}\log R}. \qquad (20)$$

In the large separation limit the logarithmic derivative of $\big\langle\delta(\boldsymbol{x}_1)\delta(\boldsymbol{x}_2)\big\rangle$ with respect to the smoothing scale is expected to be zero: this cross-correlation is simply the value of the two-point correlation function of the un-filtered field for the separation $\vert\boldsymbol{x}_1-\boldsymbol{x}_2\vert$. The resulting value for $C_{2,1}$ in the large separation limit is simply given by*

$$C_{2,1}=\frac{68}{21}+\frac{1}{3}\frac{\mathrm{d}\log\big\langle\delta^2\big\rangle}{\mathrm{d}\log R}. \qquad (21)$$

Note that this result depends on the local index of the two-point correlation function only at the smoothing scale, not at the separation at which the cells are. But this is only true for the large separation limit. The corrective term appearing in (19) for instance is likely to have a more complicated dependence with the spectrum.

The derivation of the coefficient $C_{2,2}$ can be done directly from the expression (valid in the large separation limit),

$$\begin{aligned}C_{2,2} =& \frac{16}{\big\langle\delta^2\big\rangle^2\big\langle\delta(\boldsymbol{x}_1)\delta(\boldsymbol{x}_2)\big\rangle} \\ &\times\int \mathrm{d}^3\boldsymbol{k}_1\mathrm{d}^3\boldsymbol{k}_0\mathrm{d}^3\boldsymbol{k}_2\ P(k_1)\,P(k_0)\,P(k_2) \\ &\times\exp\left[\mathrm{i}\boldsymbol{k}_0\cdot(\boldsymbol{x}_1-\boldsymbol{x}_2)\right]W(k_1\,R)\,W(k_2\,R) \\ &\times W\left[\vert\boldsymbol{k}_1+\boldsymbol{k}_0\vert\,R\right]\,W\left[\vert\boldsymbol{k}_2+\boldsymbol{k}_0\vert\,R\right] \\ &\times F_2(\boldsymbol{k}_0,\boldsymbol{k}_1)\,F_2(\boldsymbol{k}_0,\boldsymbol{k}_2)\end{aligned} \qquad (22)$$

* This result is in fact related to the value $\overline{Q}$ given by Fry (1984, eqn. 38) since one should have $C_{2,1}=2\overline{Q}$.

Using again the properties (17, 18) we get

$$\begin{aligned}C_{2,2} =& \left(\frac{68}{21}\right)^2+\frac{2}{3}\,\frac{68}{21}\frac{\mathrm{d}\log\big\langle\delta^2\big\rangle}{\mathrm{d}\log R}+\frac{1}{9}\left(\frac{\mathrm{d}\log\big\langle\delta^2\big\rangle}{\mathrm{d}\log R}\right)^2 \\ &+\frac{4}{3}\,\frac{68}{21}\int\mathrm{d}^3\boldsymbol{k}_2P(k_2)\,W(k_2\,R)\,k_2\,R\,W'(k_2\,R) \\ &\quad\times\exp[\mathrm{i}\boldsymbol{k}_2\cdot(\boldsymbol{x}_1-\boldsymbol{x}_2)]/\big\langle\delta(\boldsymbol{x}_1)\delta(\boldsymbol{x}_2)\big\rangle \\ &+\frac{1}{9}\,\frac{\mathrm{d}\log\big\langle\delta^2\big\rangle}{\mathrm{d}\log R}\int\mathrm{d}^3\boldsymbol{k}_2P(k_2)\,W(k_2\,R)\,k_2\,R\,W'(k_2\,R) \\ &\quad\times\exp[\mathrm{i}\boldsymbol{k}_2\cdot(\boldsymbol{x}_1-\boldsymbol{x}_2)]/\big\langle\delta(\boldsymbol{x}_1)\delta(\boldsymbol{x}_2)\big\rangle \\ &+\frac{1}{9}\int\mathrm{d}^3\boldsymbol{k}_2P(k_2)\,k_2^2\,R^2\,\left[W'(k_2\,R)\right]^2 \\ &\quad\times\exp[\mathrm{i}\boldsymbol{k}_2\cdot(\boldsymbol{x}_1-\boldsymbol{x}_2)]/\big\langle\delta(\boldsymbol{x}_1)\delta(\boldsymbol{x}_2)\big\rangle\end{aligned}$$

The last three terms of this expression are negligible in the large separation limit due to the presence of the factor $k_2\,R$ which, for $k_2\approx 1/\vert\boldsymbol{x}_1-\boldsymbol{x}_2\vert$, is small. The resulting value of $C_{2,2}$ is then simply given by

$$C_{2,2}=C_{2,1}^2. \qquad (23)$$

It confirms for this particular example the factorization property (15).

## 3. The generating function of the joint cumulants

The quantity we are interested in is actually the generating function $\beta(y_1,y_2)$ defined by

$$\beta_2(y_1,y_2)=\sum_{p=1,q=1}^{\infty}C_{p,q}\frac{y_1^p}{p!}\frac{y_2^q}{q!}. \qquad (24)$$

Due to the factorization property (15-16) the derivation of the $C_{p,q}$ series can then be done with the derivation of $C_{p,1}$ coefficients only, so that

$$\beta_2(y_1,y_2)=\beta(y_1)\beta(y_2) \qquad (25)$$

with

$$\beta(y)=\sum_{p=1}^{\infty}C_{p,1}\frac{y^p}{p!}. \qquad (26)$$

### *3.1. When the smoothing effects are neglected*

When one neglects the effects of the window function that is by assuming that

$W\left[\vert\boldsymbol{k}_1+\ldots+\boldsymbol{k}_p\vert\,R\right]\approx W[\boldsymbol{k}1\,R]\ldots W[\boldsymbol{k}_p\,R]$ in (12) the derivation of these integrals is quite simple. The factors $C_{p,1}$ can be seen as sums of products of vertices $\nu_p$, geometrical averages of the function $N_p$. These products correspond to tree summations for trees having one "free leg". Note that the $S_p$ parameters (Eq. [9]) on the other hand are obtained with tree summations for trees without any free leg. These tree summations have been considered by Bernardeau & Schaeffer (1992) and Bernardeau (1992). They are all related to the generating function of the vertices $\nu_p$,

$$\mathcal{G}(\tau)=\sum_{p=0}^{\infty}\nu_p\,\frac{(-\tau)^p}{p!}. \tag{27}$$

From Bernardeau (1992) the function $\mathcal{G}(\tau)$ has been shown to describe the spherical model, and is thus a known function for any cosmological model. The generating function $\beta(y)$ is then given by

$$\beta(y)=-y\frac{\mathrm{d}\mathcal{G}}{\mathrm{d}\tau}(\beta(y)). \tag{28}$$

From this implicit relationship one can easily relate $C_{2,1}$ to the value of $\nu_2=34/21$ (for an Einstein-de Sitter Universe, and one find,

$$C_{2,1}=2\nu_2=\frac{68}{21}, \tag{29}$$

which corresponds to the case (21) when $n=-3$.

### *3.2. When the smoothing effects are taking into account*

Taking into account filtering effects is of crucial importance for a practical point of view. Since the pionneering work of Goroff et al. (1986) the filtering effects are known to change the values of the $S_p$ parameters. A similar change is going to affect the coefficients $C_{p,q}$. The results of the previous sections show that it is indeed the case for the $C_{2,1}$ parameter. It is of interest to note that the relationship between $C_{p,1}$ and $S_p$ that holds in the unsmoothed case does not hold anymore. For instance we have $C_{2,1}=2/3\,S_3$ for the unsmoothed case but this is not true when the smoothing is taken into account (it would have given $68/21-2(n+3)/3$ instead of $68/21-(n+3)/3$ for $C_{2,1}$). The proper calculation of the coefficients $C_{p,1}$ should then be done with care.

#### 3.2.1. The case of two concentric cells

The basic property I use to do this calculation is to remark that the parameters $C_{p,1}$ are also involved in the expression of $\left\langle\delta^p_{R_1}(\boldsymbol{x})\delta_{R_2}(\boldsymbol{x})\right\rangle_c$, where the smoothed density fields are taken at the same position but when the scale $R_2$ is very large compared to $R_1$, $R_1\ll R_2$. This property is only true for the large separation limit. The reason is that for the calculation of such a quantity the factor $\exp[\mathrm{i}\boldsymbol{k}_0\cdot(\boldsymbol{x}_1-\boldsymbol{x}_2)]W[k_0\,R]$ in (13) has simply been replaced by $W[k_0\,R_2]$, so that this new cumulant is given by

$$\begin{aligned}\left\langle\delta^p_{R_1}\delta_{R_2}\right\rangle_c&=G_p(0,R_1)\left\langle\delta_{R_1}\delta_{R_2}\right\rangle\\&=C_{p,1}\left\langle\delta^2_{R_1}\right\rangle^{p-1}\left\langle\delta_{R_1}\delta_{R_2}\right\rangle\end{aligned} \tag{30}$$

In the following the scale $R_1$ is identified to the scale $R$ of the filtering scale and the scale $R_2$ is identified to the typical distance between cells, assumed to be larger than the smoothing scale.

#### 3.2.2. The joint PDF in Lagrangian space

The derivation of the $C_{p,1}$ parameters follows then the same scheme than the one developed for the $S_p$ parameters. The fundamental property is that the filtering in *Lagrangian* space (that is for a given mass scale) does not change the values of the cumulants. This property had been discussed for the case of a unique cell but can be easily extended to any number of concentric cells. The reason is that this property is due to angular properties of the top-hat window function similar to (17-18) (actually to the properties (B3) and (B4) of Bernardeau 1994a) that involve only the angular part of the wave vectors. As a result the filtering effects simply factorized away for concentric cells, i.e., the effect of filtering appears only in the value of $\left\langle\delta^2_R\right\rangle$ (which obvioulsy depends on $R$), of $\left\langle\delta_{R_1}\delta_{R_2}\right\rangle$, but not in the values of the $C_{p,q}$ coefficients. The derivation of the joint density PDF in different concentric cells can then be performed using the technics presented by Bernardeau & Schaeffer (1992), with the generating function of the vertices given by $\mathcal{G}(\tau)$ in Eq. (27).

So, let me now consider the joint density PDF in two different concentric cells of two different mass scales, $M_1\ll M_2$. Following the calculations made by Bernardeau & Schaeffer (1992) we get

$$\begin{aligned}p^L(\delta_1,\delta_2)=&\frac{-1}{4\pi^2}\int\frac{\mathrm{d}y_1}{\sigma^2(M_1)}\,\frac{\mathrm{d}y_2}{\sigma^2(M_2)}\\&\times\exp\left[\chi^L(y_1,y_2)+\frac{y_1\delta_1}{\sigma^2(M_1)}+\frac{y_2\delta_2}{\sigma^2(M_2)}\right]\end{aligned} \tag{31}$$

with

$$\begin{aligned}\chi^L(y_1,y_2)=&\frac{-1}{\sigma^2(M_1)}\left[y_1\mathcal{G}(\tau_1^L)-\frac{1}{2}y_1\tau_1^L\frac{\mathrm{d}}{\mathrm{d}\tau}\mathcal{G}(\tau_1^L)\right]\\&-\frac{1}{\sigma^2(M_2)}\left[y_2\mathcal{G}(\tau_2^L)-\frac{1}{2}y_2\tau_2^L\frac{\mathrm{d}}{\mathrm{d}\tau}\mathcal{G}(\tau_2^L)\right]\\\tau_1^L=&-y_1\frac{\mathrm{d}}{\mathrm{d}\tau}\mathcal{G}(\tau_1^L)-\frac{\sigma^2(M_1,M_2)}{\sigma^2(M_2)}y_2\frac{\mathrm{d}}{\mathrm{d}\tau}\mathcal{G}(\tau_2^L)\\\tau_2^L=&-y_2\frac{\mathrm{d}}{\mathrm{d}\tau}\mathcal{G}(\tau_2^L)-\frac{\sigma^2(M_1,M_2)}{\sigma^2(M_1)}y_1\frac{\mathrm{d}}{\mathrm{d}\tau}\mathcal{G}(\tau_1^L)\end{aligned} \tag{32}$$

where $\sigma(M_1)$, $\sigma(M_2)$ and $\sigma(M_1,M_2)$ are respectively the rms density fluctuations at the mass scale $M_1$, at the mass scale $M_2$ and the cross correlation between the mass scales $M_1$ and $M_2$. The superscript $^L$ designs quantities related to Lagrangian space calculations. Note that the density PDF at the mass scale $M_1$ is given by

$$\begin{aligned}p^L(\delta_1)&=\int\mathrm{d}\delta_2 p(\delta_1,\delta_2)\\&=\int\frac{\mathrm{d}y}{2\pi\mathrm{i}\sigma^2(M_1)}\exp\left[\chi^L(y_1,0)+\frac{y_1\delta_1}{\sigma^2(M_1)}\right].\end{aligned} \tag{33}$$

Our present objective is only to get the generating function of the coefficients $C_{p,1}$. It is thus necessary to limit ourself to the case where only the cumulants $\left\langle\delta_{R_1}^p\delta_{R_2}\right\rangle_c$ and $\left\langle\delta_{R_2}^2\right\rangle$ are not zero. In such a case the function $\chi^L(y_1,y_2)$ reads,

$$\chi^L(y_1,y_2)=\chi^L(y_1,0)+\frac{\sigma^2(M_1,M_2)}{\sigma^2(M_1)\sigma^2(M_2)}y_2\tau_1^L+\frac{y_2^2}{2\sigma^2(M_2)} \tag{34}$$

where $\tau_1$ is given by

$$\tau_1^L=-y_1\frac{\mathrm{d}}{\mathrm{d}\tau}\mathcal{G}(\tau_1^L). \tag{35}$$

In Lagrangian space the generating function $\beta(y)$ is simply given by

$$\beta^L(y)=\tau^L(y). \tag{36}$$

3.2.3. The joint PDF in Eulerian space

The problem is then to relate the Lagrangian density PDF to the Eulerian density PDF. The first remark of interest is that at the mass scale of $M_2$ no differences are taken into account: the reason is that the cumulants we are interested in involve only the linear order at the mass scale $M_2$. The Lagrangian mass scale $M_2$ and the Eulerian scale $R_2$ are then identical and the actual value of $\delta_2$ can be seen as a simple external constraint. We are then confronted to a problem similar to the one considered by Bernardeau (1994c) and a similar development will be followed. The relationship between the density PDF in Eulerian scale $R_1$, $p_{R_1}^E(\delta_1,\delta_2)$ and the one for Lagrangian scale $M_1$ is simply that (see Bernardeau 1994c)

$$\int_{\delta_0}^{\infty}\mathrm{d}\delta_1\ (1+\delta_1)\ p_{R_0}^E(\delta_1,\delta_2)=\int_{\delta_0}^{\infty}\mathrm{d}\delta_1\ p_{M_0}^L(\delta_1,\delta_2) \tag{37}$$

with

$$1+\delta_0=\frac{3M_0}{4\pi R_0^3\,\overline{\rho}}. \tag{38}$$

The derivation of the cumulants (in the large scale limit) can be done by identifying the exponential factor in the expression of $p^E(\delta_1,\delta_2)$. It is then interesting to note that integrating and differentiating are not going to affect the expression of the exponential term so that a simple identification can be done, taking advantage of the mass-scale relationship (38).

The calculation of the expression of $p^L(\delta_1,\delta_2)$ can be done by a saddle point approximation. The saddle point position is given by the system of equations,

$$\begin{aligned}\frac{\partial}{\partial y_1}\left[\chi^L(y_1,y_2)+\frac{\delta_1 y_1}{\sigma^2(M_1)}\right]&=0;\\ \frac{\partial}{\partial y_2}\left[\chi^L(y_1,y_2)+\frac{\delta_2 y_2}{\sigma^2(M_2)}\right]&=0.\end{aligned} \tag{39}$$

The solution, $(y_1^s,y_2^s)$, of this system is given by,

$$\begin{aligned}\delta_1&=\mathcal{G}[\tau_1^L(y_1^s)]-\frac{\sigma(M_1,M_2)}{\sigma(M_2)}y_2^s\frac{\mathrm{d}}{\mathrm{d}y}\beta(y_1^s);\\ \delta_2&=-\,y_2^s-\frac{\sigma(M_1,M_2)}{\sigma(M_1)}\beta(y_1^a).\end{aligned} \tag{40}$$

Then, at the saddle point position we have,

$$\begin{aligned}&\chi^L(y_1^s,y_2^s)+\frac{y_1^s\delta_1}{\sigma^2(M_1)}+\frac{y_2^s\delta_2}{\sigma^2(M_2)}\\ &=-\frac{\tau_1^L(y_1^0)}{2\sigma^2(M_1)}-\frac{\sigma^2(M_1,M_2)}{\sigma^2(M_1)\sigma^2(M_2)}\delta_2\beta^L(y_1^0)-\frac{\delta_2^2}{2\sigma^2(M_2)},\end{aligned} \tag{41}$$

when the expression is written up to the linear order in $\sigma^2(M_1,M_2)$. The value of $y_1^0$ is the one given by the saddle point approximation for the calculation of $p^L(M_1)$ in (33), so that,

$$\mathcal{G}[\tau_1^L(y_1^0)]=\delta_1. \tag{42}$$

In Eulerian space a similar expression is expected so that

$$\begin{aligned}&\chi^E(y_1^s,y_2^s)+\frac{y_1^s\delta_1}{\sigma^2(R_1)}+\frac{y_2^s\delta_2}{\sigma^2(R_2)}\\ &=-\frac{\tau_1^E(y_1^0)}{2\sigma^2(R_1)}-\frac{\sigma^2(R_1,R_2)}{\sigma^2(R_1)\sigma^2(R_2)}\delta_2\beta^E(y_1^0)-\frac{\delta_2^2}{2\sigma^2(R_2)}\end{aligned} \tag{43}$$

with

$$\mathcal{G}^E[\tau_1^E(y_1^0)]=\delta_1. \tag{44}$$

The expressions of $\mathcal{G}^E$, $\tau^E$ and $\beta^E$ can then simply be obtained by identification. For that the mass scales $M_1$ and $M_2$ have to be transformed into the physical scales $R_1$ and $R_2$. For the mass scale $M_2$, it is rather easy since this scale is supposed to remain linear. As a result we simply have

$$M_2=\overline{\rho}\ \frac{4\pi}{3}R_2^3. \tag{45}$$

The mass scale $M_1$ and the scale $R_1$ are related to each other with a relation similar to (38),

$$M_1=\overline{\rho}(1+\delta_1)\ \frac{4\pi}{3}R_1^3. \tag{46}$$

The identification of the various terms then gives

$$\begin{aligned}\tau^E(y_1^0)&=\tau^L(y_1^0)\ \frac{\sigma(R_1)}{\sigma\left(R_1[1+\mathcal{G}^L(\tau^L)]^{1/3}\right)};\\ \mathcal{G}^E(\tau^E)&=\mathcal{G}(\tau^L)=\mathcal{G}\left[\tau^E\frac{\sigma\left(R_1[1+\mathcal{G}(\tau^L)]^{1/3}\right)}{\sigma(R_1)}\right];\\ \beta^E(y_1^0)&=\beta_L(y_1^0)\frac{\sigma(R_1)}{\sigma\left(R_1[1+\mathcal{G}^L(\tau^L)]^{1/3}\right)}.\end{aligned} \tag{47}$$

For this result we can neglect the possible variations of $\sigma(R_1,R_2)$ with $R_1$, that is

$$\sigma\left(R_1[1+\mathcal{G}^E(\tau^E)]^{1/3},R_2\right)\approx\sigma(R_1,R_2), \tag{48}$$

since $R_1 \ll R_2$. This is a simple generalisation of the remark leading to (21) from (20) for the calculation of the coefficient $C_{2,1}$.

It is interesting to write the results in terms of the function $\tau^E(y)$,

$$\beta^E(y) = \tau^E(y) \frac{\sigma(R)}{\sigma\left(R[1+\mathcal{G}^E(\tau^E)]^{1/3}\right)}. \qquad (49)$$

The function $\beta^E(y)$ is the generating function of the coefficients $C_{p,1}$ as defined in Eq. (26). Although it has been obtained from the statistics of two concentric cells, it is identical to the one corresponding to two distant cells of the same radius. Note that, unlike the Lagrangian case, the function $\beta^E$ is no more equal to the function $\tau^E$ as a naïve extrapolation from the unsmoothed case would have given. The relation (49) is the central analytical result of this paper, and all the applications that followed are derived from this formula.

### 3.2.4. Examples of coefficients $C_{p,1}$

To start with it is possible to get, with the expansion of $\beta^E(y)$ with respect to $y$, the first coefficients $C_{p,1}$. As an illustration I give here the first six,

$$\begin{aligned}
C_{2,1} =& \frac{68}{21} + \frac{\gamma_1}{3}; \qquad (50)\\
C_{3,1} =& \frac{11710}{441} + \frac{61}{7}\gamma_1 + \frac{2}{3}\gamma_1^2 + \frac{\gamma_2}{3};\\
C_{4,1} =& 353.1 + 205.1\gamma_1 + 38.67\gamma_1^2 + 2.33\gamma_1^3 + 12.03\gamma_2\\
&+ 2.22\gamma_1\gamma_2 - 0.22\gamma_3;\\
C_{5,1} =& 6511.17 + 5497.1\gamma_1 + 1715.7\gamma_1^2 + 233.5\gamma_1^3 + 11.6\gamma_1^4\\
&+ 390.27\gamma_2 + 160.27\gamma_1\gamma_2 + 16.11\gamma_1^2\gamma_2 + 1.852\gamma_2\\
&+ 11.57\gamma_3 + 2.346\gamma_1\gamma_3 + 0.1234\gamma_4;\\
C_{6,1} =& 153708 + 171051\gamma_1 + 75460\gamma_1^2 + 16460\gamma_1^3\\
&+ 1770\gamma_1^4 + 74.7\gamma_1^5 + 13551\gamma_2 + 8878\gamma_1\gamma_2\\
&+ 1916\gamma_1^2\gamma_2 + 135.7\gamma_1^3\gamma_2 + 172.7\gamma_2^2 + 36.9\gamma_1\gamma_2^2\\
&+ 502.8\gamma_3 + 217.4\gamma_1\gamma_3 + 23.21\gamma_1^2\gamma_3 + 4.20\gamma_2\gamma_3\\
&+ 8.94\gamma_4 + 1.9\gamma_1\gamma_4 + 0.06\gamma_5;
\end{aligned}$$

with

$$\gamma_p = \frac{\mathrm{d}^p \log\left\langle\delta^2\right\rangle}{\mathrm{d}\log^p R}. \qquad (51)$$

The expression of $C_{2,1}$ have already been obtained from direct calculation in (21). I also checked that the coefficient $C_{3,1}$ can also be re-derived directly from formula (12) and using the properties (17, 18) of the top-hat window function. The results are given here for an $\Omega = 1$, $\Lambda = 0$ Universe, but, as for the $S_p$ coefficients (Bouchet et al. 1993, Bernardeau 1994b), the $\Omega$ and $\Lambda$ dependence of these coefficients are expected to be small.

### *3.3. The bias function*

The interest of such results is that it is then possible to get the expression of the joint density PDF in two distant cells. In the following I assume that the two cells have the same radius $R$. The rms density fluctuations in such cells will be denoted $\sigma(R)$, and the cross-correlation between the two cells, $\sigma^2(d)$. Moreover I will suppress the $^E$ superscript but it is always assumed. The expression of such a joint distribution function can be obtained from the form (31), with the function $\exp[\chi(y_1,y_2)]$ in (32) calculated in the large separation limit that is up to the linear order in $\sigma^2(d)$. In such a limit it reads (Bernardeau & Schaeffer 1992)

$$\begin{aligned}
\exp[\chi(y_1,y_2)] =& \left[1 + \frac{\sigma^2(d)}{\sigma^4(R)}\beta(y_1)\,\beta(y_2)\right]\\
&\times \exp\left[-\frac{\varphi(y_1)}{\sigma^2(R)} - \frac{\varphi(y_2)}{\sigma^2(R)}\right].
\end{aligned} \qquad (52)$$

As each term appearing in (52) can be factorised in

$$p(\delta_1,\delta_2) = p(\delta_1)p(\delta_2) + \sigma^2(d)\,p_b(\delta_1)\,p_b(\delta_2) \qquad (53)$$

with

$$\begin{aligned}
p_b(\delta) =& \int \frac{\mathrm{d}y}{2\pi\mathrm{i}\;\sigma^2(R)}\,\frac{-\beta(y)}{\sigma^2(R)}\\
&\times \exp\left[-\frac{\varphi(y)}{\sigma^2(R)} + \frac{\delta\,y}{\sigma^2(R)}\right].
\end{aligned} \qquad (54)$$

It is then quite convenient to define the bias function

$$b(\delta) = \frac{p_b(\delta)}{p(\delta)} = -\frac{\int \mathrm{d}y\,\beta(y)\exp\left[-\frac{\varphi(y)}{\sigma^2(R)} + \frac{\delta\,y}{\sigma^2(R)}\right]}{\int \mathrm{d}y\,\exp\left[-\frac{\varphi(y)}{\sigma^2(R)} + \frac{\delta\,y}{\sigma^2(R)}\right]}, \qquad (55)$$

which implies that the joint density PDF reads,

$$p(\delta_1,\delta_2) = p(\delta_1)\;p(\delta_2)\;\left[1 + \sigma^2(R_1,R_2)\,b(\delta_1)\,b(\delta_2)\right]. \qquad (56)$$

#### 3.3.1. Normalisation properties

In the following I will discuss the various properties of the function $b(\delta)$. The first remark is that

$$\int_{-1}^{\infty} \mathrm{d}\delta\;b(\delta)\,p(\delta) = 0 \qquad (57)$$

and

$$\int_{-1}^{\infty} \mathrm{d}\delta\;\delta\;b(\delta)\;p(\delta) = 1. \qquad (58)$$

In Fig. 5 I present the shape of the bias function for a particular set of parameters. The general features that are seen is that it is a growing function of $\delta$. It is (roughly) of the sign of $\delta$. Note that for a purely Gaussian field, one would have

$$b_{\mathrm{Gaussian}}(\delta) = \frac{\delta}{\sigma^2}. \qquad (59)$$

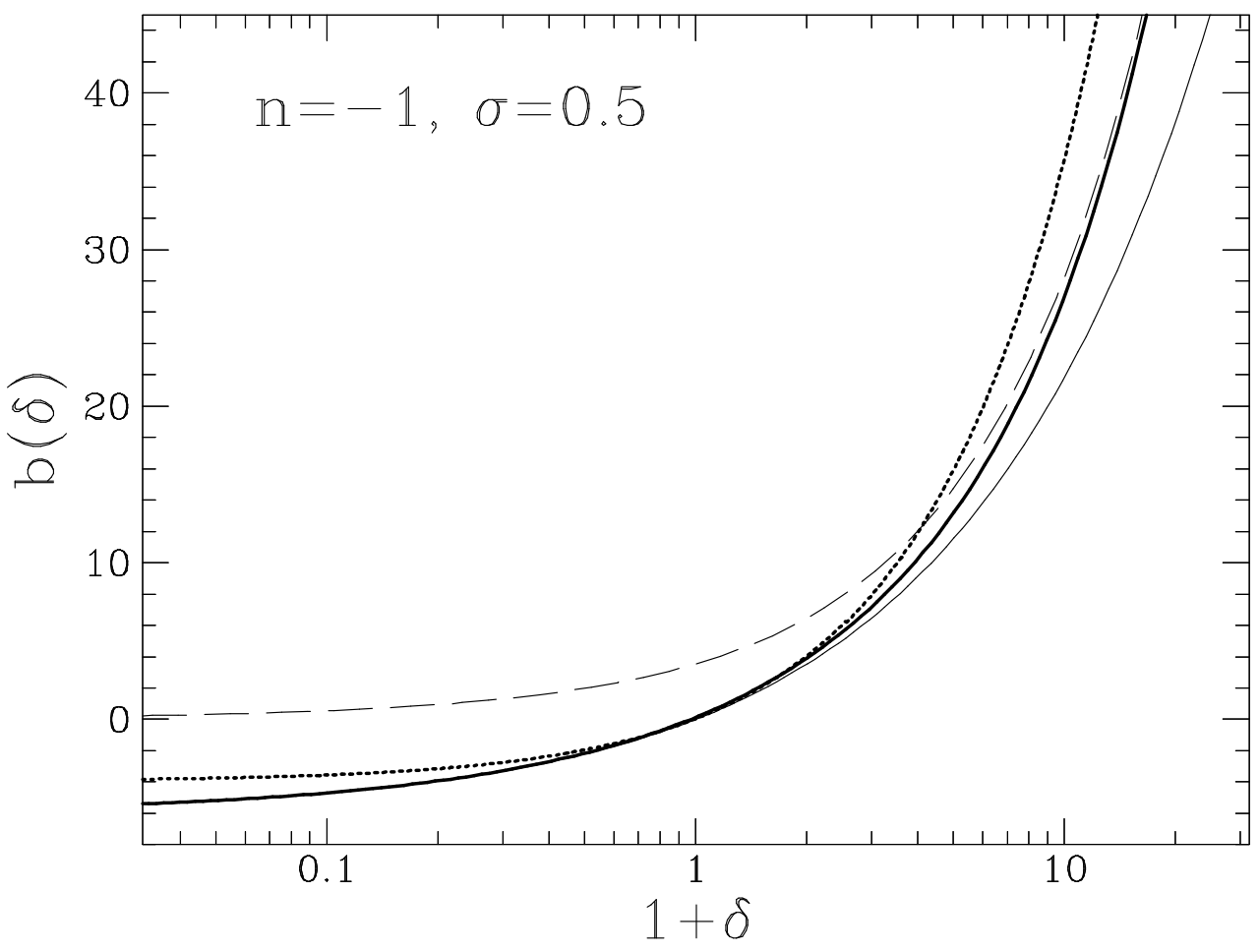

**Fig. 4.** The approximate form (64) (thin solid line) compared to the exact integration of (55) (thick solid line). The thin dashed line is the form (65) and the thick dotted line is the Gaussian case (59).

This curve is shown as a thick dotted line.

3.3.2. Analytic approximate forms

It is useful to remember that the function $\mathcal{G}(\tau)$ in (47,49) can be fruitfully approximated by

$$\mathcal{G}(\tau)=\left(1+\frac{2\tau}{3}\right)^{-3/2}-1. \tag{60}$$

In the following I will also assume that the power spectrum can be approximated by a power law behaviour so that

$$P(k)\propto k^{n}. \tag{61}$$

It implies that

$$\mathcal{G}^{E}(\tau)=\mathcal{G}\left(\tau\left[1+\mathcal{G}^{E}(\tau)\right]^{-(n+3)/6}\right), \tag{62}$$

and

$$\beta^{E}=\frac{3}{2}(1+\mathcal{G}^{E})^{(n+3)/3}\left[(1+\mathcal{G}^{E})^{-2/3}-1\right]. \tag{63}$$

An analytical approximate formula for $b(\delta)$ can be obtained with the saddle point approximation. It implies that

$$\begin{aligned}b(\delta)=&-\frac{\beta^{E}(y^{0})}{\sigma^{2}(R)}\\ =&\frac{-3}{2\sigma^{2}(R)}(1+\delta)^{(n+3)/3}\left[(1+\delta)^{-2/3}-1\right].\end{aligned} \tag{64}$$

In fig. 4 I show this approximate form compared to an exact integration of the expression (55). the approximation is very good for $\delta\lesssim 0$ but becomes quite inaccurate when the density contrast is large.

Another possible approximation is the following approximation,

$$\begin{aligned}b(\delta)=-\int{\rm d}y\ \beta^{E}(y)\exp[(1+\delta)/\sigma^{2}]\\ \Big/\int{\rm d}y\ \varphi(y)\exp[(1+\delta)/\sigma^{2}],\end{aligned} \tag{65}$$

which is a function of $(1+\delta)/\sigma^2$ only. As shown in Figs. 4 and 7, this approximation is valid when the value of $\delta$ is large, and the larger $\sigma$ is the better it is.

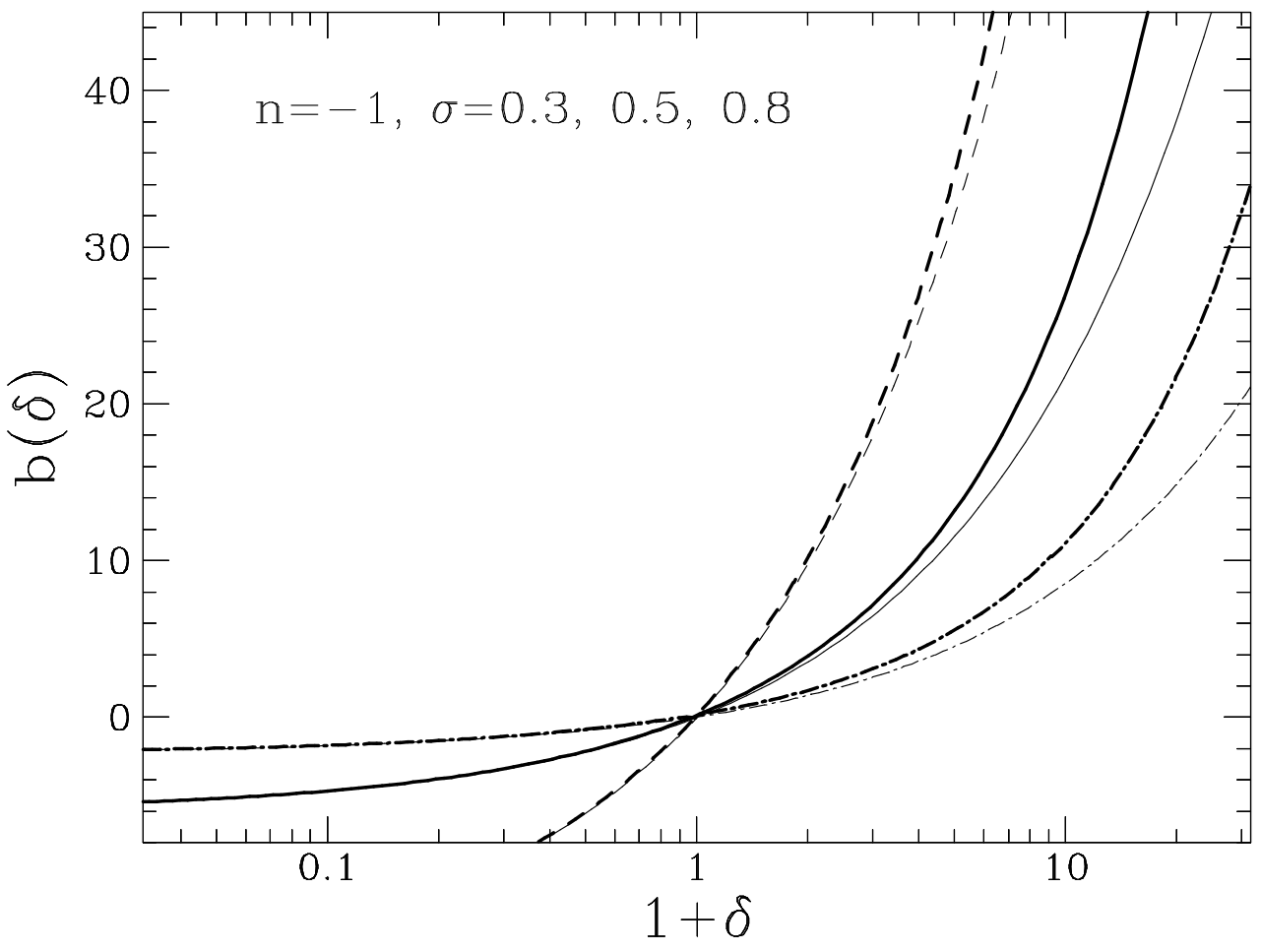

**Fig. 5.** The variation of the bias function for different values of $\sigma$. The approximate form (64) (thin lines) are compared to the exact integrations (55) (thick lines). The dashed line, solid line and dot dashed line are respectively for $\sigma=0.3$, $\sigma=0.5$ and $\sigma=0.8$.

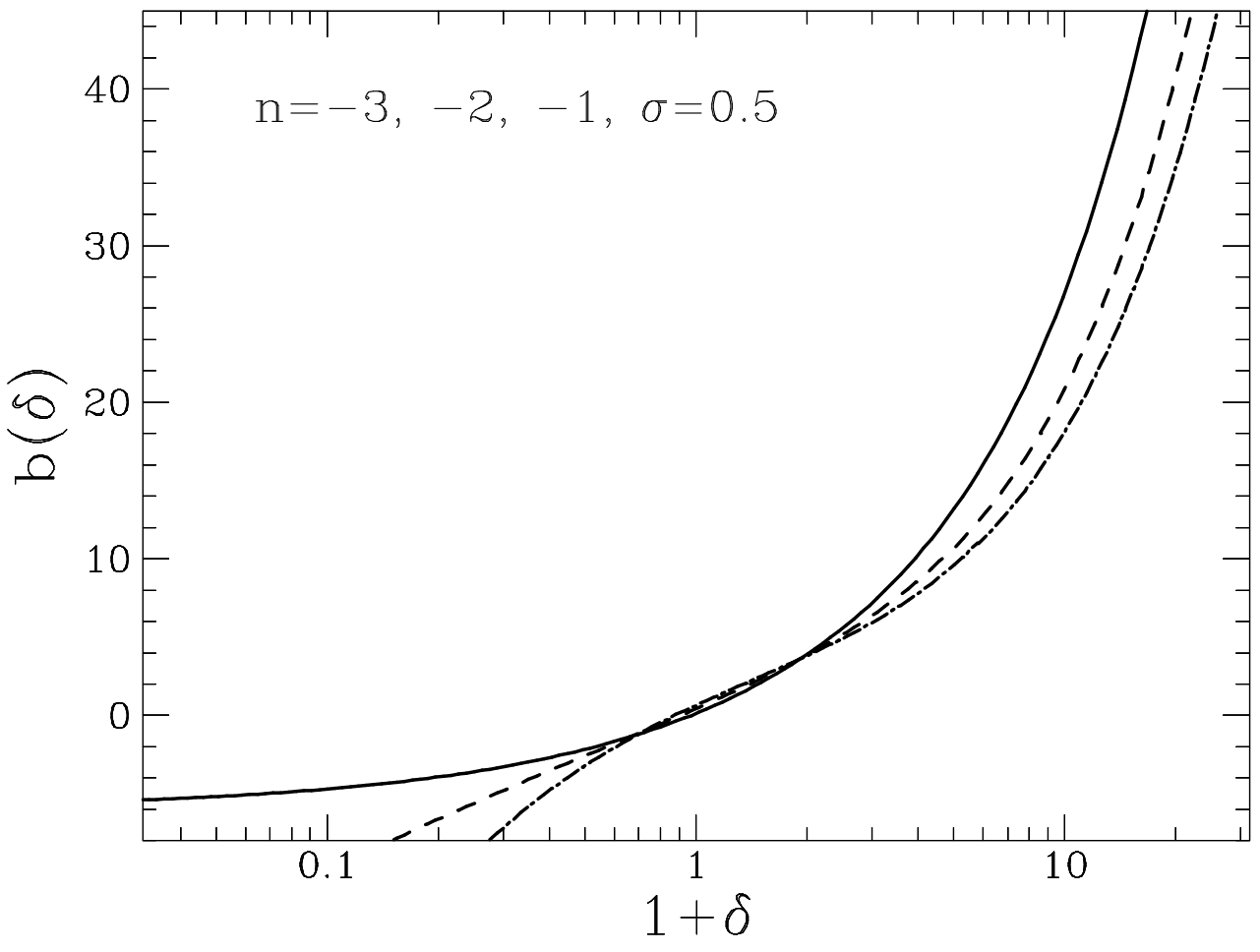

**Fig. 6.** The variation of the bias function for different values of the power law index $n$. The solid line, dashed line and dot dashed line are respectively for $n=-1$, $n=-2$, $n=-3$

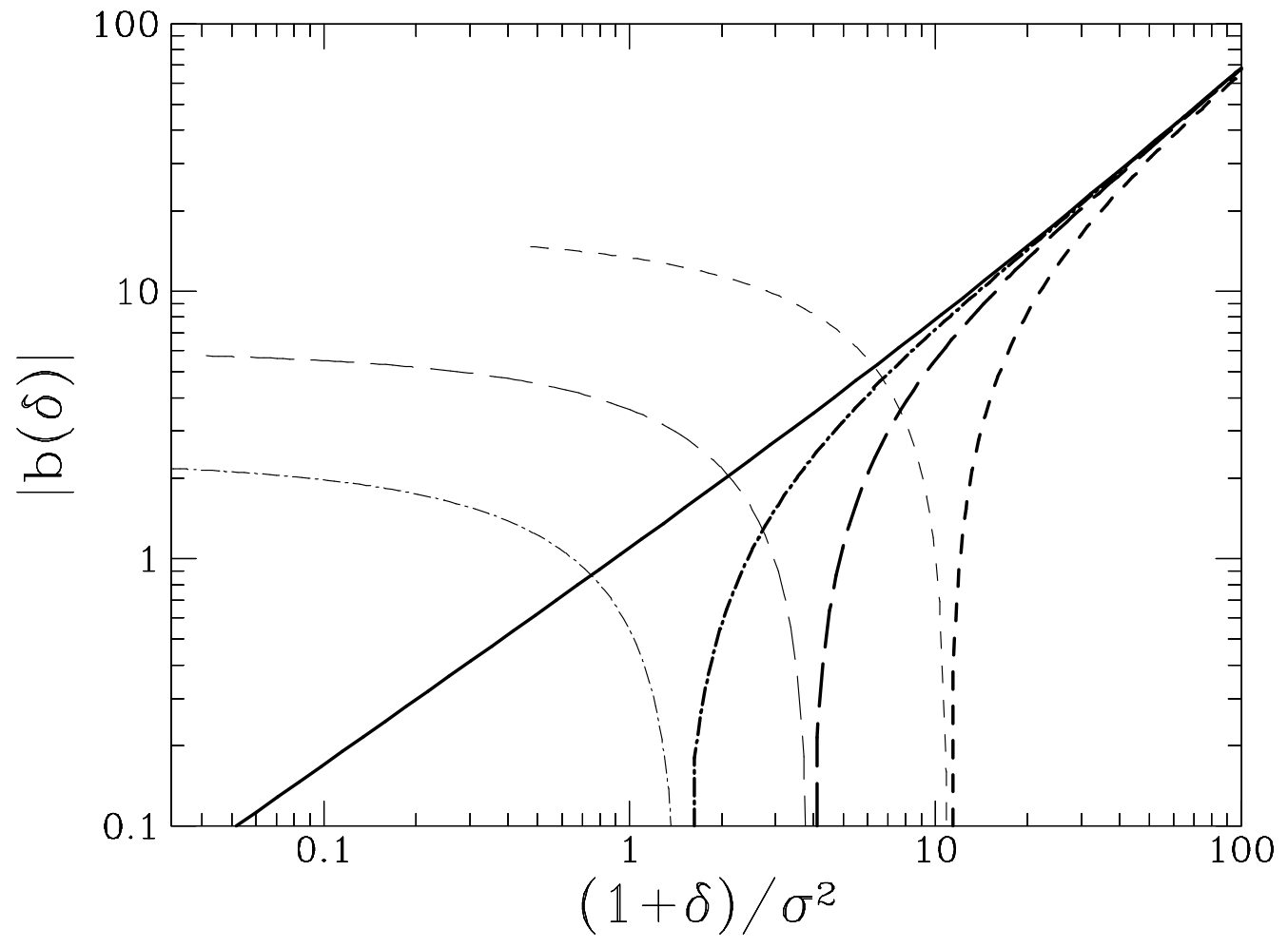

**Fig. 7.** The bias function expressed as a function of $(1+\delta)/\sigma^2$ for different values of $\sigma$. The solid line is the approximation (65). The dashed line, long dashed line and dot dashed line are respectively for $\sigma=0.3$, $\sigma=0.5$ and $\sigma=0.8$ and the thin lines correspond to negative values of $b(\delta)$.

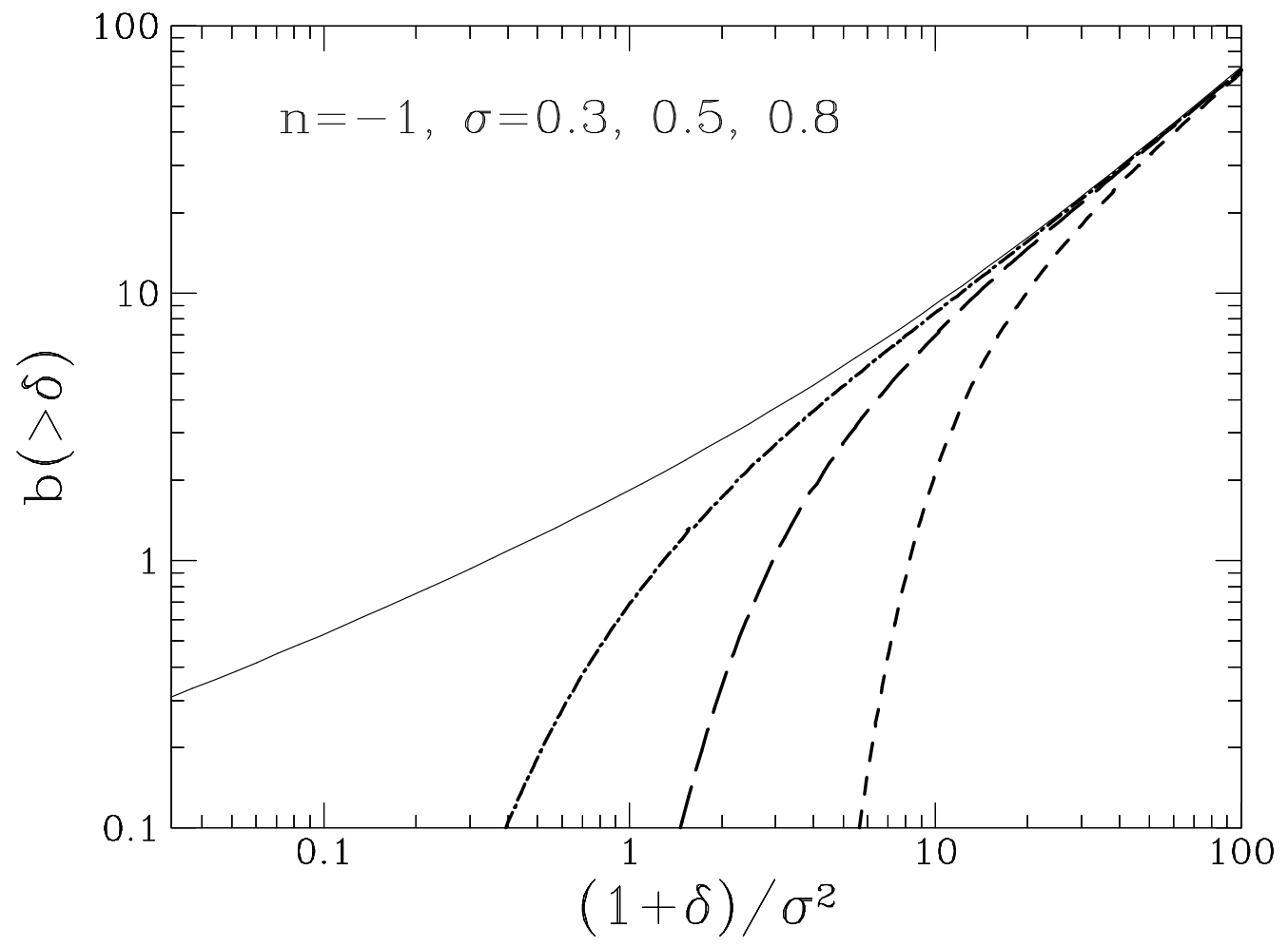

**Fig. 8.** The cumulative bias function (66) expressed as a function of $(1+\delta)/\sigma^2$ for different values of $\sigma$. The symbols are the same than in Fig. 7.

In Figs. 5 and 6 I present the shape of the function $b(\delta)$ for different values of the index $n$ and of $\sigma$. the variation of the shape of $b(\delta)$ for different values of $n$ are important for negative values of $\delta$. For positive values of $\delta$ the variations are not so important. The approximation (64) is however quite inadequate to see this latter property. The approximation (65) would have been more accurate for this regime.

The Fig. 7 is quite instructive to show that the $\sigma$ variations, for large values of $\delta$, are well described by the approximation (65). The function $b(\delta)$ indeed appears to be dependent only on $(1+\delta)/\sigma^2$.

A quantity that I will consider for the comparison with numerical simulation is the bias function for a threshold. It is defined by

$$b(>\delta_0)=\frac{\int_{\delta_0}^{\infty}\mathrm{d}\delta\ b(\delta)\ p(\delta)}{\int_{\delta_0}^{\infty}\mathrm{d}\delta\ \ p(\delta)}. \tag{66}$$

In Fig. 8 I give the shape of the function $b(>\delta)$ expressed as a function of $(1+\delta)/\sigma^2$ and for different values of $\sigma$.

### 3.3.3. Asymptotic behaviour

When $\delta$ is large the form (54,55) naturally reduces to the form (65). The integrals when $\delta$ is large are dominated by the singular point $y_s$ of the expression of $\varphi(y)$,

$$\varphi(y)-\varphi_s\approx -a_s\ (y-y_s)^{3/2}. \tag{67}$$

At this point the function $\beta^E(y)$ reads

$$\beta^E(y)-\beta^E_s\approx -b_s\ (y-y_s)^{1/2}. \tag{68}$$

The asymptotic behaviour can then be calculated following the method of Bernardeau & Schaeffer (1992) and Bernardeau (1994c). We obtain

$$b(\delta)\approx b_*\ \frac{\delta}{\sigma^2}, \tag{69}$$

with

$$b_*=\frac{3}{2}\frac{b_s}{a_s}. \tag{70}$$

Peculiar values of these parameters are given in table 1 for different values of $n$.

**Table 1.** The parameters of the critical point as a function of the spectral index, $n$, for the bias function.

| $n$ | $a_s$ | $b_s$ | $b_*$ |
|---|---|---|---|
| $-3$ | 1.84 | 0.528 | 0.431 |
| $-2.5$ | 2.21 | 0.644 | 0.438 |
| $-2$ | 2.81 | 0.836 | 0.447 |
| $-1.5$ | 3.93 | 1.206 | 0.460 |
| $-1$ | 6.68 | 2.154 | 0.484 |

The resulting value of $b_*$ is slightly lower than 0.5 and is only weakly dependent of $n$. It is also interesting to note that this asymptotic behaviour gives a lower value of the bias than for the Gaussian case for a given value of $\delta/\sigma^2$. Note, however, that in the present paper the value of $\delta$ is not the initial overdensity of the fluctuation but its final value. The two appraoches, peak correlation in the initial Gaussian density field and correlation in the final quasi-linear field, then cannot be compared directly.

## 4. Validity of the large separation approximation

The validity of the large separation limit is quite difficult to test in general. The reason is that a lot of terms have been neglected to get the results (50) for the large separation limit. With numerical integrations it is however possible to estimate the value of $C_{2,1}$ for any separation. To do the numerical calculations I use the shape of the 3-point correlation function obtained from the second-order perturbation theory given by Fry (1984),

$$\xi_3(\boldsymbol{r}_1,\boldsymbol{r}_2,\boldsymbol{r}_3) = \left[\frac{10}{7} - \frac{3+n}{n}\,\cos(\theta_1)\,\left(\frac{r_{21}}{r_{31}}+\frac{r_{31}}{r_{21}}\right) + \frac{4}{7}\,\frac{-3-2n+(n+3)^2\cos^2\theta_1}{n^2}\right]\xi_2(\boldsymbol{r}_1,\boldsymbol{r}_2)\,\xi_2(\boldsymbol{r}_1,\boldsymbol{r}_3) + \mathrm{cyc.}(1,2,3) \quad (71)$$

where $\theta_1$ is the angle between the vectors $\boldsymbol{r}_2-\boldsymbol{r}_1$ and $\boldsymbol{r}_3-\boldsymbol{r}_1$ and $r_{21}$ and $r_{31}$ are the lengths of these vectors. Two points are then assumed to be in a cell of radius $R$ at the position $\boldsymbol{x}_0$ and the other in another cell of radius $R$ at the position $\boldsymbol{x}_1$ with $|\boldsymbol{x}_0-\boldsymbol{x}_1| = d$. The resulting coefficient, $C_{2,1}$, is then a function of $d/R$ and $n$ only. In Fig. 9, I give the results for power law spectra of index $n=-1.3$ and $n=-1$.

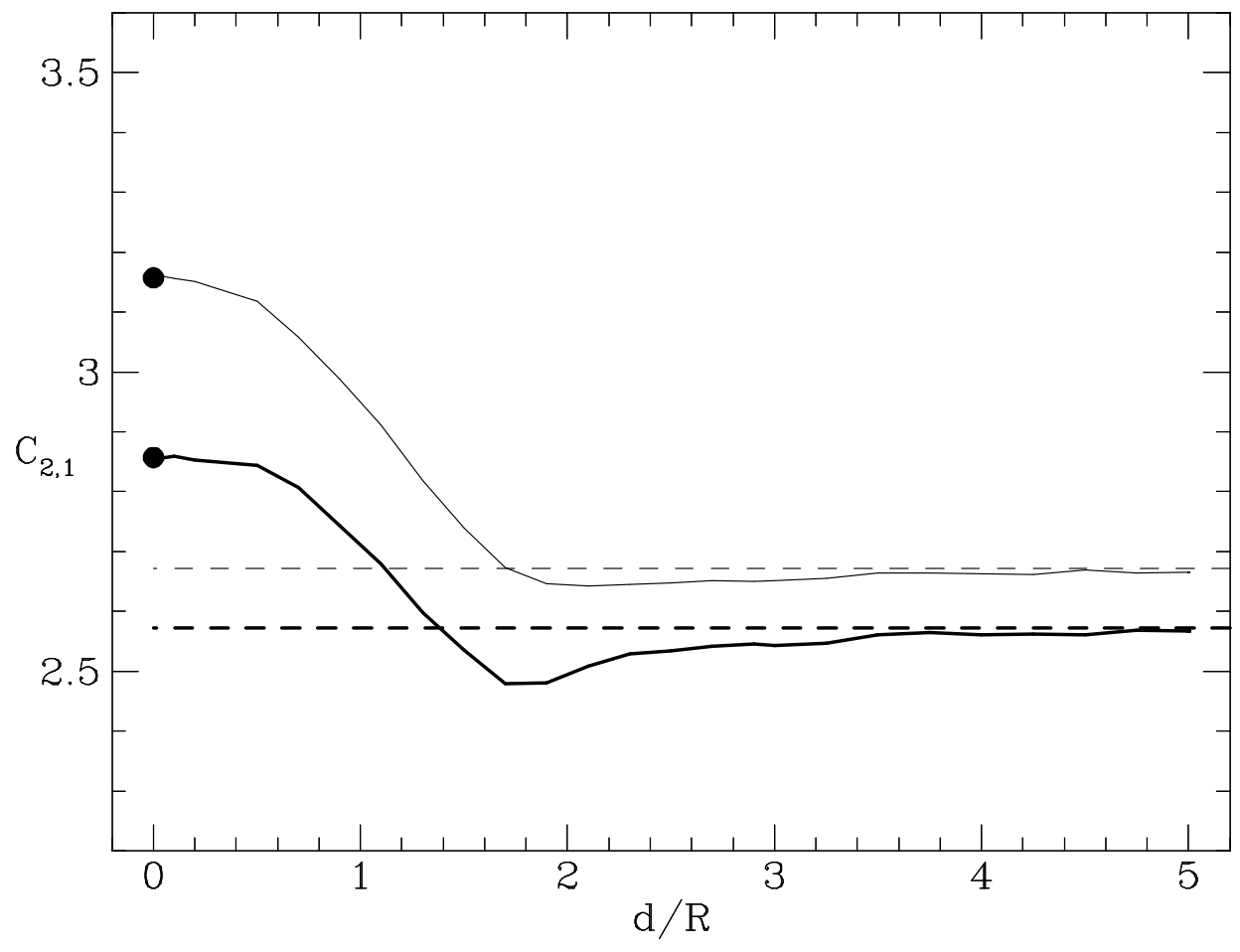

**Fig. 9.** Numerical evaluation of the $C_{2,1}$ coefficient as a function of separation $d$. The calculations have been made for $n=-1.3$ (thin lines) and $n=-1$ (thick lines). The circle is the $d=0$ prediction (in such a case $C_{2,1}(d)=S_3$) and the dashed line is the large separation limit (21).

When $d=0$ the two cells are exactly superposed so that the coefficient $C_{2,1}$ is exactly $S_3$ (Eq. [9]). For $d/R \gg 1$ one expects the value of $C_{2,1}$ to converge to the large separation limit (21). And it can be seen that this limit is indeed obtained very rapidly, as soon as $d/R \gtrsim 3$. Actually the various corrective terms (appearing in (19) and (20), tend to cancel each other. It is a bit surprising however that $C_{2,1}$ is not a monotonic function of the distance. It has a minimum for $d \approx 2R$, that is when the cells are side by side. Actually although the values of $S_3$ and $C_{2,1}$ in the large separation limit depend only on the local index at the smoothing scale the whole shape the function $C_{2,1}$ as a function of the distance may depend on other characteristics of the power spectrum.

## 5. Comparison with numerical simulations

For comparison with numerical results, I use the numerical simulation done by Couchman (1991) with CDM initial conditions for $h=0.5$. This simulation uses an adaptive $\mathrm{P^3M}$ code with $2.1\ 10^6$ particles. The simulation was made in a cubic box of $200h^{-1}$Mpc size with periodic boundary conditions. I used the simulation at a time-step for which the linearly extrapolated rms density fluctuation in a spherical box of $8h^{-1}$Mpc radius is unity.

### *5.1. The measure of the coefficient $C_{2,1}$*

For the determination of the various statistical properties, I measured the number of particles in $200^3$ spherical cells disposed on a regular grid. Each sphere is then associated by pairs to 6 neighboring spheres that are at the distance $d$ along each direction of the grid. By summing over the particle content of each sphere (appropriately weighted) multiplied by the content of its neighboring spheres, it is possible to estimate the value of statistical quantities such as the coefficients $C_{p,q}$. However, an accurate measurement of merely the coefficient $C_{2,1}$ turns out to be quite difficult to obtain even for rather small separation. For instance I found

$$C_{2,1} = 1.9 \pm 1.2 \quad (72)$$

for $R = 10h^{-1}$Mpc and $d = 20h^{-1}$Mpc where the error-bar has been obtained by dividing the simulation in four subsamples. As it can be noticed the estimated errors are pretty large, making the comparison with the theoretical prediction rather un-conclusive. A precise analysis of this problem is beyond the scope of this paper. Note that the possible sources of errors for the one-point density PDF have been investigated in details by Colombi et al. (1995). It turns out the such coefficients are very sensitive to finite size effects, a problem which might be amplified here by the fact that I considered pair of cells. It may also be due to the fact that the number of pairs that have been used for the determination of joint moments is slightly too small, making the sampling of the large density tail of the distribution too poor for an accurate determination of the moments. These problems led me to focus my analysis on the behaviour of the global bias function.

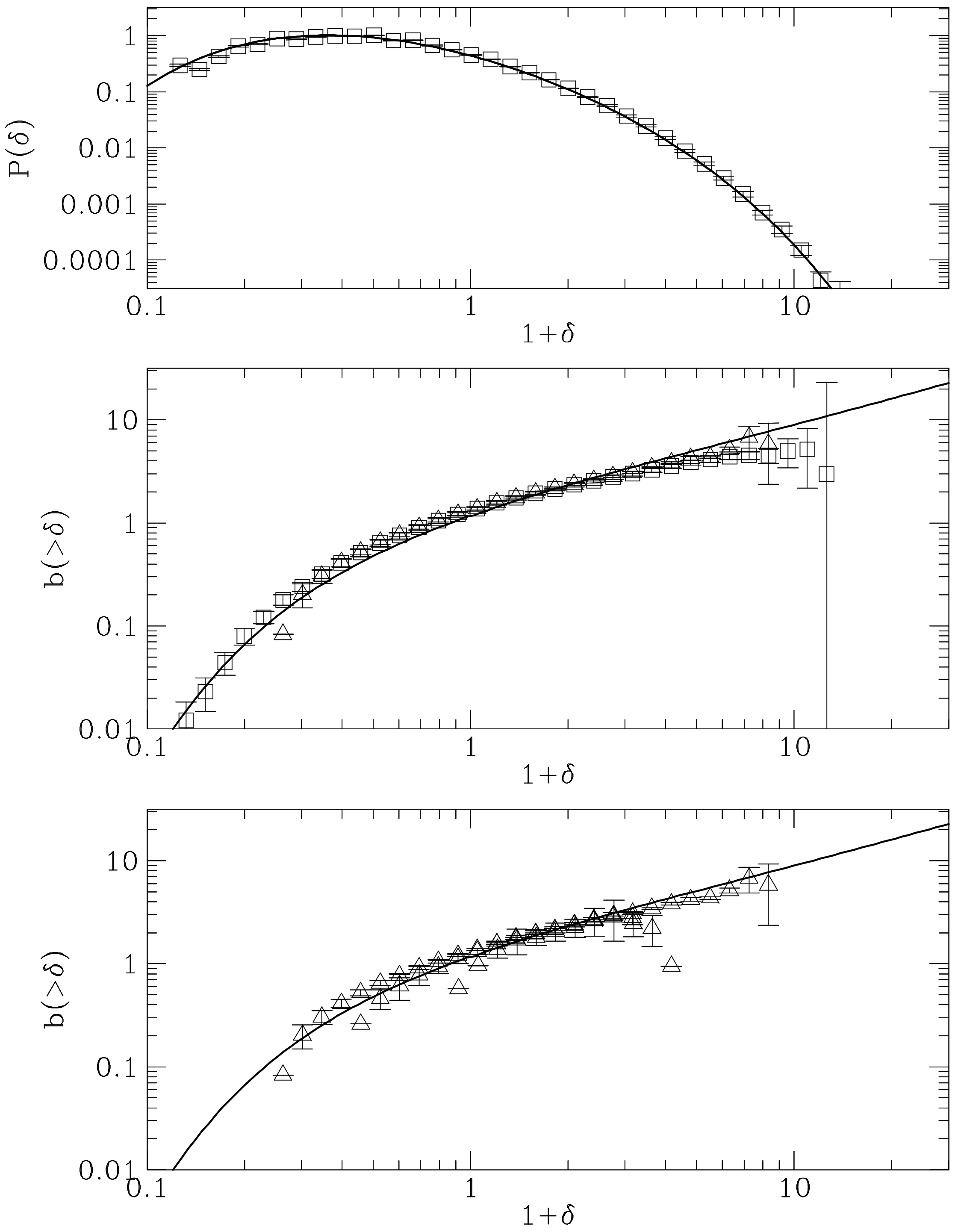

**Fig. 10.** The measured bias function for $R = 5h^{-1}$Mpc in a numerical simulation with CDM initial conditions. For this scale $\sigma = 0.95$ and $n = -1.3$. The upper panel shows the measured density distribution $p(\delta)$ compared to the predictions from Perturbation Theory (solid line). The central panel shows the measured function $b(>\delta)$ with the formula (73) (squares) and with the formula (77) (triangles). The distance between the centers of the cells is $d = 10h^{-1}$Mpc. The bottom panel shows the function $b(>\delta)$ obtained with the formula (77) when the distance between the centers of the cells varies from $d = 10h^{-1}$Mpc to $d = 20h^{-1}$Mpc. The error-bars have been obtained by the measure of these quantities in four different subparts of the simulation. The solid lines are the prediction of Perturbation Theory for the large separation limit, Eq. (55).

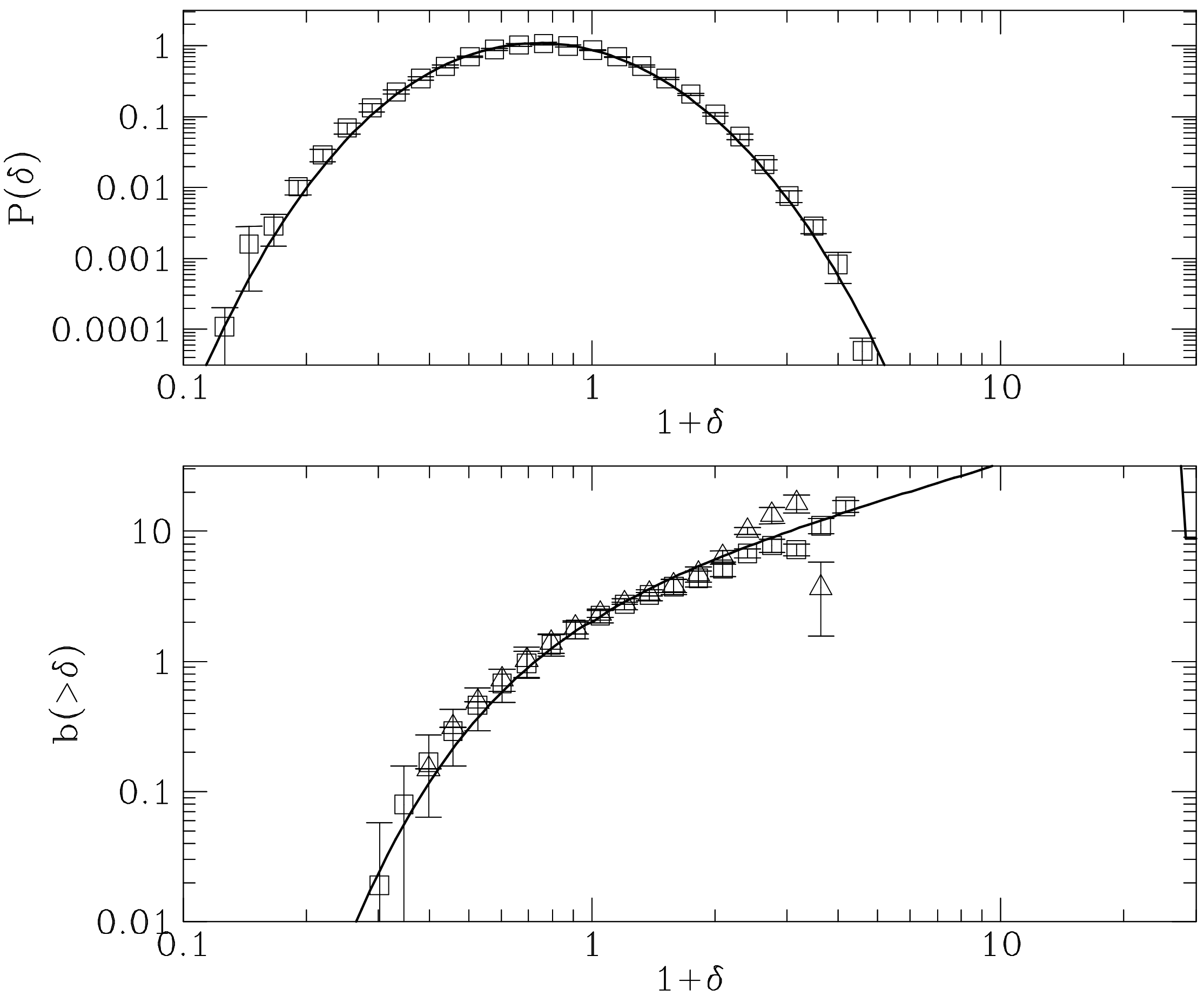

**Fig. 11.** The measured bias function for $R = 10h^{-1}$Mpc. For this scale $\sigma = 0.46$ and $n = -1$ and the distance between the cells is $d = 20h^{-1}$Mpc. The symbols are the same as in Fig. (10)

*5.2. The measure of the bias function*

The direct measure of the bias function turns out to be easier to do in practice. I considered two ways to get the bias function. In the first method I estimated

$$b_1(>\delta_0) = \frac{1}{\langle\delta_1\delta_2\rangle} \times \left[\frac{\int_{\delta_0}^{\infty}\mathrm{d}\delta_1\int_0^{\infty}\mathrm{d}\delta_2\,(1+\delta_2)\,p(\delta_1,\delta_2)}{p(>\delta_0)} - 1\right] \quad (73)$$

where

$$\langle\delta_1\delta_2\rangle = \int_0^{\infty}\mathrm{d}\delta_1\ \mathrm{d}\delta_2\ (1+\delta_1)(1+\delta_2)p(\delta_1,\delta_2) - 1. \quad (74)$$

In practice the quantity $p(>\delta_0)$ is estimated by determining the fraction of cells in which the number of particles is greater than $N_0 \equiv (1+\delta_0)\overline{N}$ ($\overline{N}$ is the mean number of particles in a cell). Then the integrals intervening in the expression of $b_1$ are estimated by

$$\int_{\delta_0}^{\infty}\mathrm{d}\delta_1\int_0^{\infty}\mathrm{d}\delta_2\,(1+\delta_2)\,p(\delta_1,\delta_2) = \frac{1}{200^3}\sum_{\text{cells }N(i)>N_0}\frac{N_2(i)}{6}; \quad (75)$$

and

$$\int_0^{\infty}\mathrm{d}\delta_1\int_0^{\infty}\mathrm{d}\delta_2\,(1+\delta_1)\,(1+\delta_2)\,p(\delta_1,\delta_2) = \frac{1}{200^3}\sum_{\text{cells}}N(i)\,\frac{N_2(i)}{6}, \quad (76)$$

where $N(i)$ is the number of particles in cell $i$ and $N_2(i)$ is the total number of particles in the six neighboring cells at the distance $d$.

The second method is based on the determination of,

$$b_2(>\delta_0) = \frac{1}{\langle\delta_1\delta_2\rangle}\left[\frac{\int_{\delta_0}^{\infty}\mathrm{d}\delta_1\ \mathrm{d}\delta_2 p(\delta_1,\delta_2)}{[p(>\delta_0)]^2} - 1\right]^{1/2}. \quad (77)$$

This expression is obtained by counting the fraction of pairs of cells for which the number of particles exceeds a fixed number $N_0$ in both cells.

The main interest in considering these two approaches is that for the form (56) the two functions should coincide. It is thus a test for the factorization property (56).

The numerical results are given in Figs. 10 and 11. The comparison of the bias functions (73) and (77) (respectively the squares and the triangles) show that they coincide nicely (within the error-bars). This is an important result since it means that the form (56) for the joint density PDF is very robust and holds even when the large separation limit is not reached. It is of interest however to have in mind that the properties (15) and (56) are equivalent only in the large separation limit. So this numerical result does not mean at all that the property (15) holds for any separation.

The dependence of the bias function with the separation is presented in Fig. 10. For the distance at which the bias function can be reliably measured it is found to be very stable and independent of the distance. It checks another aspect of the form (56), that is that the dependence with the distance $d$ enters only through the value of the mean two point correlation function between the two cells. It implies in other words that the bias is very weakly dependent of the distance.

The Figs. 10 and 11 also show that the theoretical prediction for the bias function gives a very good description of the measured shape of $b$. This is quite remarkable since the measures are not made at a very large separation.

## 6. Discussion

A reliable description of joint matter density PDF-s has been obtained based on PT. The analytical results are valid in principle in the large separation limit and for a smoothing scale larger than the correlation length. Comparison with results of numerical simulations proved that they provide us with a good description of the joint density PDF for a broad range of parameter. In particular they are still accurate when the rms density fluctuation approaches unity at the smoothing scale and when the distance equals only two times the distance of the smoothing length.

Remarkable properties have been obtained using PT in the large separation limit. In particular the property (15) of the $C_{p,q}$ coefficients leads to the factorization form (56). The latter property has been confirmed by the numerical analysis and is shown to be in fact a good description of the joint density PDF for any value of the distance (larger than $2R$). This implies that the dependence with the distance enters only through the cross-correlation between the cells, $\sigma^2(d)$, and that the dependence of the bias factor with the two densities can be factorized in $b(\delta_1)\,b(\delta_2)$. It is however worth reminding that the extension of the factorization property (56) to small distances does not prove that the property (15) also holds for small distances.

Moreover, the analytic work gives the expression of the bias function $b(\delta)$ from the large separation limit approximation, and these analytical results are in good agreement with the numerical measurements. It suggests that the gravitationally induced bias does not vanish at large scale and is well described by the quasi-linear dynamics.

The accuracy of the theoretical prediction is very encouraging and shows that the theoretical expression of the bias can indeed be used for quantitative predictions.

Its implication for the estimation of the error-bars on statistical measurements related to the one-point density PDF due to finite sample effects will be exploited in a coming paper.

For direct astrophysical applications, (e.g. the determination of what could be the bias associated with the observed clusters) the analytic results obtained in this paper have two important limitations;

- contrary to the calculations of Bardeen et al. (1986) this is not the correlation of peaks. However it is expected that, in the rare event limit, the results are not affected: a very dense spot in the density field is indeed expected to be a peak.
- The other limitation is that the calculations have been made for a large smoothing length, larger than the size of the clusters. This property is due to the use of PT calculations and can hardly be overcome except with pure numerical analysis.

It anyway indicates that gravity is expected to induce specific biases, that are not only due to peak selection effects as for the results of Bardeen et al. (1986). Note that we anyway do not expect qualitative changes from this previous picture: the bias factor associated with the clusters is expected to be independent of scale and proportional to its mass in the large richness limit (eq. 69). Such properties were also predicted to be a generic result for the correlation of dense spots in a strongly nonlinear cosmological field by Bernardeau & Schaeffer (1992). I leave for another paper the estimation of more quantitative predictions.

*Acknowledgements.* I am very grateful to Hugh Couchman for providing the results of the N-body simulation, and to Richard Schaeffer for useful discussions.